\documentstyle[aps,multicol,prl,epsfig,floats]{revtex}

\begin{document}

\twocolumn[\hsize\textwidth\columnwidth\hsize\csname@twocolumnfalse%
\endcsname

\title{Evidence for Kosterlitz-Thouless type orientational ordering of \\
CF$_3$Br monolayers physisorbed on graphite}

\author{S. Fa{\ss}bender, M. Enderle, K. Knorr}
\address{Technische Physik, Universit\"at des Saarlandes,
66041 Saarbr\"{u}cken, Germany}

\author{J. D. Noh and H. Rieger}
\address{Theoretische Physik, Universit\"at des Saarlandes,
66041 Saarbr\"{u}cken, Germany}
\draft

\maketitle

\begin{abstract}
  Monolayers of the halomethane CF$_3$Br adsorbed on graphite have been
  investigated by x-ray diffraction. The layers crystallize in a
  commensurate triangular lattice. On cooling they approach a
  three-sublattice antiferroelectric pattern of the in-plane
  components of the dipole moments. The ordering is not consistent
  with a conventional phase transition, but points to
  Kosterlitz-Thouless behavior. It is argued that the transition 
  is described by a 6-state clock model on a triangular lattice with
  antiferromagnetic nearest neighbor interactions which is studied with
  Monte-Carlo simulations. A finite-size scaling analysis shows that
  the ordering transition is indeed in the KT universality class. 
\end{abstract}

]

\newcommand{\A}{\AA}

Two-dimensional systems cannot show long-range order by breaking
a continuous symmetry at any finite temperature $T$. Nevertheless
Kosterlitz and Thouless (KT) demonstrated that the 2D XY-ferromagnet
with nearest neighbor interaction and equivalently the planar rotator
exhibit a phase transition from an ordered phase with an algebraic
decay of the spin correlations to a disordered phase via the unbinding
of vortex-antivortex pairs \cite{KT}.

The KT-concept has been applied successfully to the onset of
superfluidity \cite{bishop,rudnick} and of superconductivity
\cite{beasley} in thin films and two-dimensional melting
\cite{nelson}. An application to layered magnetic systems with large
ratios of the intra- and the inter-layer exchange appears to be more
direct, but here the KT transition is masked by the transition to 3D
behavior \cite{regnault}. In the present article we present x-ray
diffraction data on CF$_3$Br monolayers physisorbed on graphite and
present evidence that the orientational {\it pseudospin} ordering is
of the KT-type. To our knowledge the present study reports the first
observation of a KT-transition in a (quasi)-magnetic system.

Physisorbed monolayers are indeed good approximations to
two-dimensional (2D) systems, as evidenced by the fact that phase
transitions occurring in such layers can be often described by the 2D
versions of elementary models of statistical mechanics such as the
Ising or the 3-state Potts model \cite{beretz,feng,fassbender}.

The halomethane CF$_3$Br is a prolate C$_{3v}$ molecule with a dipole
moment of about 0.5~D.  Monolayers have been adsorbed on exfoliated
graphite and investigated by x-ray powder diffraction. The coverage
$\rho$, temperature $T$ phase diagram is rather complex
\cite{maus,knorr1}.  Here we concentrate on a coverage which is
representative of the extended monolayer regime in which the monolayer
lattice is commensurate with the graphite lattice. Diffraction
patterns are shown in Fig.~1. The peaks of the patterns have been
analyzed for position, intensity, and the coherence length, as has been
described elsewhere~\cite{fassbender,knorr1}. The commensurate layer
melts around 105~K~\cite{knorr2}. In the 2D solid state, the
diffraction pattern is dominated by the principal Bragg reflection of
the commensurate, triangular $2\times2$ lattice at
Q$_p$=1.475 \A$^{-1}$. Just below the melting temperature, this peak is
in fact the only one visible.  Down to lowest $T$, the peak position is
independent of $T$, as imposed by the commensurability with the
substrate. The peak width yields a coherence length of about 
180~\A\ throughout the 2D solid state which compares well with the average
lateral size of the crystallites of exfoliated graphite usually quoted
in the literature.  The $2\times2$ structure has been also observed
for monolayers of CF$_4$ \cite{croset}, CF$_3$Cl \cite{weimer},
C$_2$F$_6$ \cite{arndt} on graphite. These molecules and CF$_3$Br have
the threefold symmetry and the CF$_3$ group in common. Hence it is
plausible to assume that CF$_3$Br stands on the substrate, presumably
with the F$_3$ tripod down. An isolated CF$_3$Br would prefer to lie
flat on the substrate, but definitely the $2\times2$ mesh is too tight
to accommodate the molecules in this orientation. On the other hand
steric considerations suggest that the lateral $2\times2$ packing of
perpendicular molecules can tolerate tilt angles of the molecular axis
up to 30$^\circ$ with respect to the substrate normal. A tilt is equivalent
to a non-zero in-plane component of the dipole moment. We regard this
component as pseudospin $S_i$. In this sense the $2\times2$ state is
disordered with a zero time average of every $S_i$, and is stabilized
at higher $T$ by a libration and/or a precession of the molecular axis
about the substrate normal.

\begin{figure}[t]
\centerline{\epsfig{file=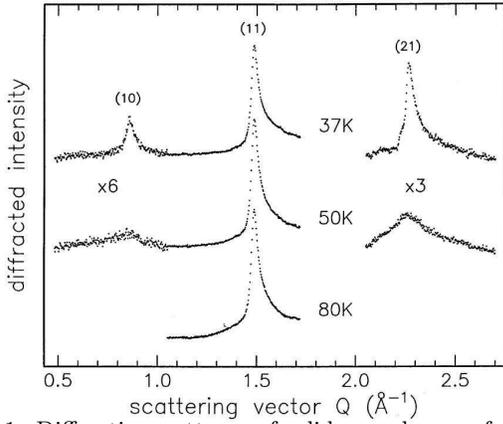, width=0.8\columnwidth}}
\caption{Diffraction patterns of solid monolayers of CF$_3$Br on graphite 
  ($\rho=1.08$). The (hk) indices refer to the $2\sqrt{3}\times2\sqrt{3}$
  superlattice. The blind region is due to the strong (002) reflection
  of the graphite substrate.}
\label{fig1}
\end{figure}

On cooling, two additional features develop in the diffraction pattern
which finally, below 40~K, can be regarded as Bragg peaks at
Q=0.851~\A$^{-1}$ and 2.253~\A$^{-1}$. They are easily identified as
superlattice reflections brought forth by a tripling of the 2D mesh
leading to a $2\sqrt{3}\times2\sqrt{3}$ commensurate hexagonal lattice.
The (hk) indices of Fig.~1 refer to this superlattice. The (20)
reflection, which should occur at 1.703~\A$^{-1}$, is absent. Below 40~K,
the coherence length derived from the width of the superlattice
reflections is again 180~\A\ and is thus imposed by the substrate. The
global structure of the low temperature state must be consistent with
the hexagonal plane group p3 with three molecules in the supercell.
Regarding the molecules as rigid, there are five positional
parameters, namely the translational coordinates x,y and the three
Eulerian angles of one molecule. The low number of reflections does
not allow a rigorous determination of these parameters, nevertheless
one can make some semi-quantitative statements. A perpendicular
orientation of the molecules in combination with a staggered azimuthal
angle of the CF$_3$ group about the substrate normal is not sufficient
to explain the superlattice intensities. The cell tripling must
involve tilts of the molecular axes away from the perpendicular
direction, presumably in combination with slight displacements of the
molecule centers out of the original positions. A satisfactory
agreement of measured and calculated intensities (including the
accidental absence of the (20) reflection) is obtained with tilts of
about 20$^\circ$ pointing toward a next neighbor such that the pseudospins
form an antiferro-``magnetic'' 120$^\circ$ Ne\'el-pattern on a 2D triangular
lattice.

\begin{figure}[t]
\centerline{\epsfig{file=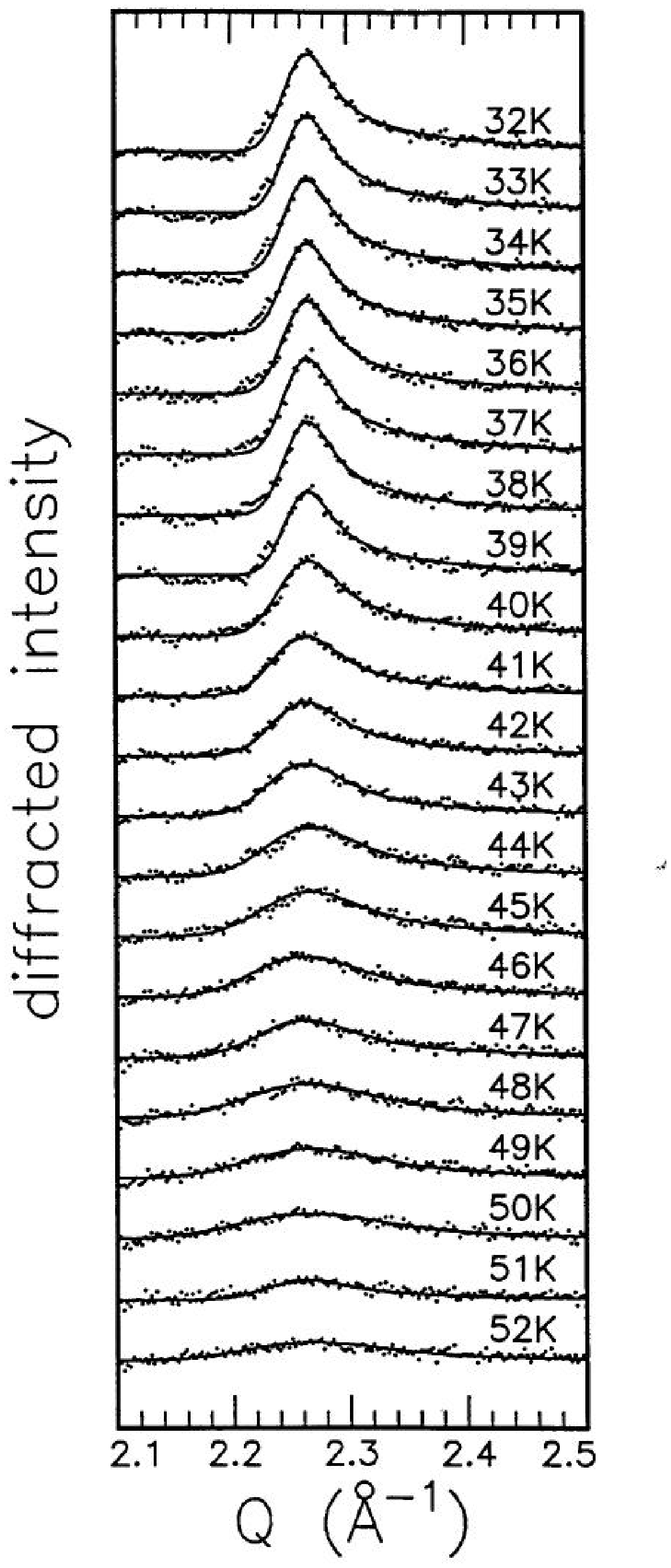, width=0.65\columnwidth}}
\caption{The temperature evolution of the (21) superlattice reflection.}
\label{fig2}
\end{figure}

In case of a conventional order-disorder transition, the integrated
intensities of the superlattice reflections would be a measure of the
order parameter and should hence vanish when approaching the critical
temperature from below. (See ref. \cite{fassbender} for such a
transition for C$_2$F$_5$Cl on graphite). This is however not what is
observed in the experiment on CF$_3$Br monolayers. On heating the
superlattice peaks broaden while keeping approximately their integral
intensity until they are finally lost in the background. The
temperature evolution of the (21) reflection is shown in Fig.~2
together with fits of a theoretical profile. The $T$-dependence of the
coherence length $\xi$ as determined from the intrinsic width of this
peak is shown in Fig.~3.  The solid line of this figure is a fit of the
KT-expression 
\begin{equation}
\xi=A\exp(B(T/T_K-1)^{-1/2})
\label{KT-form}
\end{equation}
to the data for $T>40$~K. The fit parameters are $A=9\pm2\;$\A,
$B=1.5\pm0.4$, $T_K=30\pm3\;$K. Note that the value of $A$ is reasonably
close to the lattice parameter of the 2D mesh. 
Thus $\xi$ is
expected to diverge at a KT-critical temperature $T_K$ of about 30~K,
but the growth of the correlated regions is interrupted when $\xi$ reaches
the size of the graphite crystallites. This happens at about 40~K.

\begin{figure}[t]
\centerline{\epsfig{file=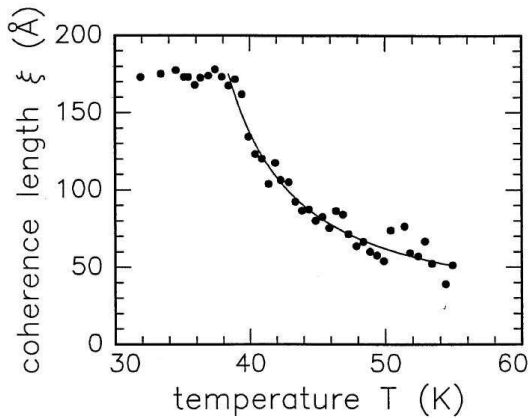, width=0.8\columnwidth}}
\caption{The temperature dependence of the coherence length of the 
  antiferroelectric correlations, as derived from the intrinsic width
  of the (21) reflection.}
\label{fig3}
\end{figure}

Given the scatter of the data points of Fig.~3, it is clear that fits
with power laws $\xi\propto(T/T_C-1)^{-\nu}$ are likewise possible. We
do think, however, that an interpretation of the intensity between 40~K
and 55~K in terms of conventional critical diffuse scattering, $T>T_C$,
is not meaningful, because of the absence of any critical $T$-dependence
of the integrated intensities, and because the diffuse scattering
would be presumably far below the detection limit in x-ray or neutron
diffraction studies on adsorbed monolayers on exfoliated graphite. At
least we do not know of any successful experiment of this type.

The heat capacity measurement also shows an indication of the phase
transition~\cite{knorr2}. Apart from the melting anomaly,
the heat capacity of commensurate CF$_{3}$Br monolayers shows
an anomaly extending from 35~K to 40~K which is just the $T$-range 
where $\xi$ saturates. However it does not allow a quantitative analysis
of the critical scaling behavior since the anomaly is very weak.

Do CF$_3$Br monolayers meet the theoretical requirements for a
KT-transition? Clearly the pseudospin correlations are bound to a
plane, thus the system is 2D with respect to the relevant degrees of
freedom at the phase transition. The pseudospin is presumably not a
strictly isotropic planar rotator but experiences a crystal field
which breaks the continuous azimuthal symmetry. It has been shown
however that --- at least for the planar XY ferromagnet --- the
KT-behavior is stable with respect to small crystal fields of 6-fold
symmetry \cite{jose}. The site symmetry of the monolayers is indeed
6-fold, but the ordered structure approached is not ferro- but
antiferroelectric.  Antiferro-type ordering on a triangular lattice is
affected by frustration with the consequence that the helicity, a
discrete 2-fold symmetry, enters into the problem. The order-disorder
transition is then described by a confluence of the Ising and KT
universality classes \cite{lee,others} with an Ising type anomaly of
the specific heat, but the spin correlations of the disordered phase
still follow the KT-form of $\xi(T)$ as we will demonstrate now.

\begin{figure}
\centerline{\epsfig{file=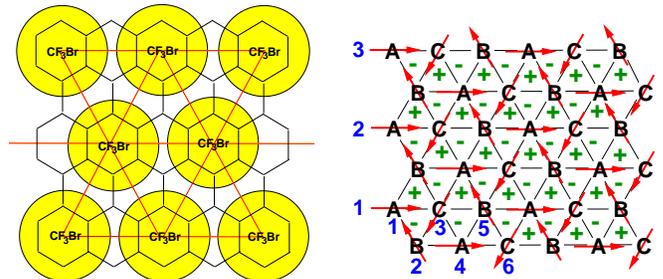,width=\columnwidth}}
\caption{{\bf Left:} The spatial arrangement of the CF$_3$Br molecules 
  on the underlying hexagonal graphite surface in the $2\times2$
  structure that is present below 105K. The resulting trigonal
  lattice is also shown. {\bf Right:} A triangular lattice of size
  $12\times 3$ with the 6-state clock spins indicated as arrows
  representing the 6 preferred directions of the CF$_3$B dipolar
  moments.  The arrow configuration shown is one of the 12 ground
  states by which the lattice sites are separated into three $A$, $B$,
  and $C$ sub-lattices. The helicity for each elementary triangle
  corresponding to this ground state is also indicated by $+$ and $-$.}
\label{fig4}
\end{figure}

From the above considerations follows that the pseudo-spins formed by
the CF$_3$Br dipoles are arranged on a triangular lattice, interact
antiferromagnetically with nearest neighbors \cite{remark} and have 6
preferred orientations. Hence the model that captures the universal
features of the orientational ordering transition should be the
anti-ferromagnetic six-state clock model on a 2D $N = L_x\times L_y$
triangular lattice~(see Fig.~\ref{fig4}).  The six-state clock spin is
a planar spin pointing toward discrete six directions; ${\bf S} =
(\cos\theta,\sin\theta)$ with $\theta = \frac{2\pi n}{6}$ ($n =
0,1,\ldots,5$).
The Hamiltonian of this model reads
\begin{equation}
{\cal H} = 2 J \sum_{\langle i,j\rangle} \cos(\theta_i - \theta_j) \ ,
\label{ham}
\end{equation}
where the sum is over nearest neighbor site pairs $\langle i,j\rangle$
and $J>0$ is the antiferromagnetic coupling strength.  The overall
factor $2$ is introduced for a computational convenience. It is a
limiting case of an antiferromagnetic $XY$ model with infinite
anisotropy field $\cos{p\theta}$ ~($p=6$).

The antiferromagnetic coupling on a triangular lattice induces a
frustration. As a result the system has 12-fold degenerate ground
states with three-sublattice structure. Global spin rotations by
multiples of $2\pi/6$ and also Ising like transformation that inverts
all helicities formed by the spins on elementary triangles (see Fig.~4)
transforms one ground state into another. Therefore the model
possesses a $C_6~(\mbox{6-fold clock}) \times Z_2~(\mbox{Ising})$
symmetry and we expect that there are two phase transitions associated
with the spontaneous breaking of the $C_6$ and $Z_2$ symmetry.

We performed extensive Monte-Carlo simulations of the model
(\ref{ham}) using conventional single spin-flip dynamics with up to
$10^8$ Monte-Carlo steps per spin and lattice sizes up to
384$\times$192 and studied its phase transition with a thorough finite
size scaling analysis. Here we focus on the breaking of the $C_6$
symmetry that is indicated by the x-ray diffraction experiments
reported above, the details of the breaking of the $Z_2$ symmetry
(which happens at a slightly higher temperature and turns out to be in
the Ising universality class) will be reported elsewhere \cite{long}.

The order parameter for the $C_6$ symmetry is the
sublattice magnetization
\begin{equation}
m_A = \frac{1}{N} \sum_{i\in A} \exp(i\theta_i) \ ,
\end{equation}
where the sum runs over only sites in the $A$ sublattice ($m_B$ and
$m_C$ can be defined similarly). Analogously the spatial spin
correlation function is defined as
\begin{equation}
C_m (r) = \left \langle \frac{3}{N} \sum_{i\in A} \cos( \theta_i -
\theta_{i+r}) \right\rangle
\label{corr}
\end{equation}
and the susceptibility is given by $\chi_m=N\langle m_A^2\rangle$.

\begin{figure}[t]
\centerline{\epsfig{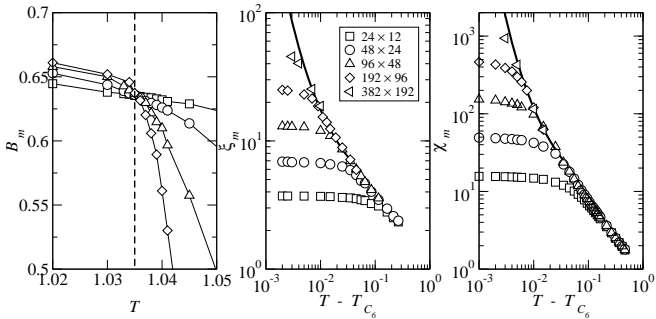}}
\caption{{\bf Left:} The Binder parameter $B_m$ (\ref{binder}) 
  versus temperature for different system sizes. The intersection
  point of the different curves locates the critical temperature
  indicated by the broken line at $T_{C_6}=1.035 \pm 0.0005$. {\bf
    Middle:} The correlation length (\ref{xi-from-C}) as a function of
  $T-T_{C_6}$ (with $T_{C_6}=1.035$) in a log-log plot. Upwards
  curvature of the numerical data indicate already a
  faster-than-algebraic divergence. The full line is a least
  square-fit to the KT-form (\ref{KT-form}). {\bf Right:} The magnetic
  susceptibility with a fit to the expected KT-form.  }
\label{fig5}
\end{figure}

The transition temperature is estimated using the dimensionless 
ratio of moments (or Binder parameter)
\begin{equation}
B_m = 1 - \frac{\langle m_A^4\rangle}{3 \langle m_A^2 \rangle^2} \;,
\label{binder}
\end{equation}
which is shown in Fig.~5 and yields the critical temperature
$T_{C_6} = 1.035 \pm 0.0005$. We extract the correlation length 
from the correlation function $C(r)$ defined in (\ref{corr}) via
\begin{equation}
\xi_m^2 = \frac{ \sum_r r^2 C_m(r)}{\sum_r C_m(r)}.
\label{xi-from-C}
\end{equation}
Our data are shown in Fig.~5 in a log-log plot of $\xi_m(T)$ versus
$T-T_{C_6}$ with $T_{C_6}$ from above. They fit nicely to the KT-form
(\ref{KT-form}) with $A=2.54$ and $B=0.193$. Note that the
non-universal number, $B$, differs significantly from the one for the
experimental data, which is not unusual for microscopically different
systems in the KT universality class (see e.g.\ \cite{pierson}). The
susceptibility, also shown in Fig.~5, follows the expected KT-form
$\chi_m=A'\exp(B'(T-T_{C_6})^{-1/2})$, too. We also checked other
quantities (like the decay of $C_m(r)$ at $T_{C_6}$) and found
everything to be consistent with a KT-scenario. 

In summary, our measurements show that CF$_3$Br molecules physisorbed
on graphite are arranged in a commensurate triangular lattice. On
cooling, a N\'eel-type 120$^\circ$ pattern is approached. The growth of the
critical correlations has been followed over a wide temperature range.
It finally stops when the correlation length reaches the size of the
substrate crystallites. The behavior is consistent with the
KT-scenario. An antiferromagnetic 6-state clock model on a triangular
lattice, which we studied numerically, describes the universal
features of this ordering transition well.

This work has been supported by the Deutsche Forschungsgemeinschaft
(project Kn234/9 and SFB 277).


\begin{references}
\bibitem{remark} The character of the relevant orientation dependent
  interactions of CF$_3$Br is not known, but a comparison of
  monolayers of several polar methane derivatives \cite{knorr1}
  suggests that for CF$_3$Br the short range anisotropic part of the
  intermolecular van der Waals force and hard core repulsion are more
  important than the medium range dipole-dipole interaction.
\end{references}
\end{document}